\documentclass[12pt]{iopart}
\usepackage{graphics,graphicx,dcolumn,bm,epic,eepic,float,fleqn}
\usepackage[latin1]{inputenc}
\usepackage[dvips]{epsfig}
\usepackage[square,numbers,compress]{natbib}

\graphicspath{{}{images/}}

\def\epsilon{\varepsilon}
\def\theta{\vartheta}

\def\rho{\varrho}

\begin{document}
\title{Simulations of slip flow on nanobubble-laden surfaces}

\author{J.~Hyv\"aluoma$^{1}$, C.~Kunert$^{2}$, and J.~Harting$^{3,2}$}
\address{
$^{1}$ MTT Agrifood Research Finland, FI-31600 Jokioinen, Finland\\
$^{2}$ Institute for Computational Physics, University of Stuttgart, Pfaffenwaldring 27, D-70569
Stuttgart, Germany\\
$^{3}$ Department of Applied Physics, Eindhoven University of Technology, Den Dolech 2, NL-5600MB Eindhoven,
The Netherlands\\
}
\ead{j.harting@tue.nl}

\date{\today}

\begin{abstract}
On microstructured hydrophobic surfaces, geometrical patterns may lead to
the appearance of a superhydrophobic state, where gas bubbles at the
surface can have a strong impact on the fluid flow along such surfaces.
In particular, they can strongly influence a detected slip at the surface.
We present two-phase lattice Boltzmann simulations of a flow over
structured surfaces with attached gas bubbles and demonstrate how the
detected slip depends on the pattern geometry, the bulk pressure, or the
shear rate. Since a large slip leads to reduced friction, our results
allow to assist in the optimization of microchannel flows for large throughput.
\end{abstract}

\submitto{\JPCM}

\maketitle

\section{Introduction}
The impressive success of the miniaturization of technical devices down to
submicrometric sizes has led to numerous practically applicable devices
which are used for the manipulation and transport fluids.  Such
microfluidic devices include micromixers, DNA amplifiers, chromatography
systems, or chemical reactors. However, in order to fully understand the
properties of such devices and to optimise them for efficient usage, some
fundamental questions in physics including the behavior of single
molecules in fluid flow or the validity of the no-slip
boundary condition have to be
answered~\cite{bib:tabeling-book,bib:lauga-brenner-stone-05}. While for
gas flows at large Knudsen number rarefaction effects that
can result in a slippage of the flow at surfaces are expectecd, 
for liquids one would
naively assume that its velocity close to a surface always corresponds to
the actual velocity of the surface itself. This assumption is called the
no-slip boundary condition and can be accounted for as one of the
generally accepted fundamental concepts of fluid mechanics. However, this
concept was not always well accepted. Some centuries ago there were long
debates about the velocity of a Newtonian liquid close to a surface and
the acceptance of the no-slip boundary condition was mostly due to the
fact that no experimental violations could be found, i.e., a so-called
boundary slip could not be detected. 

Recently, well controlled experiments have shown a violation of the
no-slip boundary condition in sub-micron sized geometries. Since then,
mostly
experimental~\cite{bib:lauga-brenner-stone-05,craig-neto-01,bib:tretheway-meinhart-04,cheng-giordano-02,choi-westin-breuer-03,baudry-charlaix-01,bib:cottin-bizone-etall-02,vinogradova-yakubov-03,bib:tretheway-meinhart-04}
but also theoretical works~\cite{vinogradova-95,bib:degennes-02}, as well
as computer
simulations~\cite{succi02,bib:barrat-bocquet-99,bib:cieplak-koplik-banavar-01,bib:thompson-troian-1997,bib:tretheway-zhu-petzold-meinhart-2002}
have been performed to improve our understanding of slippage.  An
(apparent) violation of the no-slip boundary condition can be explained by
the complex behavior of a fluid close to a solid interface which
involves the interplay of many physical and chemical properties. These
include the wettability of the solid, the shear rate or flow velocity, the
bulk pressure, the surface charge, the surface roughness, as well as
impurities, dissolved gas, or bubbles attached to surfaces. Due to the
large number of different parameters, a significant dispersion of the
results can be observed for almost similar
systems~\cite{bib:lauga-brenner-stone-05,neto-etal-05}. For example,
observed slip lengths vary between a few
nanometres~\cite{bib:churaev-sobolev-somov-84} and
micrometres~\cite{bib:tretheway-meinhart-04} and while some authors find a
dependence of the slip on the flow
velocity~\cite{choi-westin-breuer-03,bib:zhu-granick-01,craig-neto-01},
others do not~\cite{cheng-giordano-02,bib:tretheway-meinhart-04}.
Extensive reviews of the slip phenomenon have recently been published by
Lauga et al.~\cite{bib:lauga-brenner-stone-05}, Neto et
al.~\cite{neto-etal-05}, as well as Bocquet and
Barrat~\cite{bib:bocquet-barrat-2007}.

A boundary slip is typically quantified by the so-called slip length
$b$ -- a concept that was already proposed by Navier in 1823. He
introduced a boundary condition where the fluid velocity at a surface is
proportional to the shear rate at the surface~\cite{bib:Navier} (at
$x=x_0$), i.e., 
\begin{equation}
u_z(x_0)=b\frac{\partial u_z(x)}{\partial x}. 
\end{equation}
In other words, the slip length $b$ can be defined as the distance
from the surface where the flow velocity vanishes.

During the last few years, the substantial scientific research invested in
the slip phenomenon has lead to a more clear picture which can be
summarized as follows: one can argue that many surprising results
published were only due to artefacts or misinterpretation of experiments.
In general, there seems to be an agreement within the community that the
no-slip boundary condition is a valid assumption for smooth hydrophilic
surfaces down to
contact~\cite{vinogradova-yakubov-03,bib:cottin-bizone-etall-05,vinogradova.oi:2009}.
Slip lengths larger than a few nanometres can usually be referred to as
``apparent slip'' and are often caused by experimental artefacts or not
fully understood effects. Examples for those can be surface structures,
variations in the local contact angle at the surface or dissolved gases
forming nano- or microscale bubbles on the surface. It is also possible
to utilize such surface properties in order to generate bubbles in a
controlled way so as to add slippery surfaces to the flow channel and
thereby increase the slip length and reduce the viscous friction.  An
example that has lately gathered a lot of interest are surfaces that are
patterned with arrays of holes where gas bubbles are attached to the
holes~\cite{bib:steinberger-cottin-bizonne-kleimann-charlaix:2007,bib:jens-jari:2008,bib:davis-lauga-2009,bib:gao-feng:2009,Teo10,tsai-etal-09}. Although
reasonably flat bubbles lead to increased effective slip lengths, it has
recently been observed that bubbles strongly protruding to the flow
channel may actually have an opposite effect, i.e., they may reduce the
slip and increase the hydrodynamic
drag~\cite{bib:steinberger-cottin-bizonne-kleimann-charlaix:2007}. Numerical
studies have predicted even a possibility of negative slip
lengths~\cite{bib:steinberger-cottin-bizonne-kleimann-charlaix:2007,bib:jens-jari:2008,Teo10}. The enhanced friction can
be understood with help of the additional roughness produced by the
bubbles which can overcome the lubricating effect of the stress-free
liquid-gas interfaces. The numerical results of negative slip lengths were
recently verified in theoretical investigation of Davis and
Lauga~\cite{bib:davis-lauga-2009}.

In all studies concerning slip flow over a bubble mattress we are aware of, the mattress
under consideration has been built up of evenly sized and equally shaped
bubbles. However, as real configurations inevitably include some kind of
imperfections, it would also be interesting to know whether the disorder
in the mattress affects the detected slip behavior. Therefore, in this paper
a case where the mattress is made up of bubbles of two different sizes
is considered. In practice, such a
situation is achieved by structuring the surface with holes that have
different radii. This structuring leads, for a given bulk pressure, to a
bubble configuration where bubbles located in larger holes protrude
stronger to the flow channel than those in the smaller ones. In
particular, we show that the bubble-size distribution can significantly
affect the slip behavior and, e.g., decrease the negative slip caused by
strongly protruding bubbles. While this reduction may hamper experimental
observations of negative slip, it may on the other hand be utilized in
designing bubble surfaces aiming to maximal flow slippage.

\section{Lattice Boltzmann simulations to investigate boundary slip}
The simulation method used to study microfluidic devices has to be chosen
carefully. 
Molecular
dynamics simulations (MD) are the best choice to simulate the fluid-wall
interaction, but the presently available computer power is not sufficient to simulate
length and time scales necessary to achieve orders of magnitude which
are relevant for experiments. However, boundary slip with a slip length
$b$ of the order of many molecular diameters has been studied
with molecular dynamics simulations by various 
authors~\cite{baudry-charlaix-01,bib:cieplak-koplik-banavar-01,bib:thompson-troian-1997,bib:cottin-bizone-etall-04,bib:priezjev-darhuber-troian-2005}.

The current contribution focuses on numerical investigations of the slip
phenomenon by means of lattice Boltzmann simulations.  It should be
noticed that while a large number of groups utilizes the lattice Boltzmann
technique to investigate microfluidic problems, only a very small number
of researchers is actually applying the method to study slippage. This is
surprising since mesoscopic simulation methods offer a closer relation to
experimentally relevant time and length scales than microscopic techniques
such as molecular dynamics, even though the interactions between fluids
and surfaces have to be described on a mesoscopic scale in the lattice
Boltzmann method.

Within the lattice Boltzmann method a popular approach is to introduce slip by generalizing
the no-slip bounce-back boundary conditions in order to allow specular
reflections with a given
probability~\cite{succi02,bib:tretheway-zhu-petzold-meinhart-2002,bib:tang-tao-he-2005,bib:sbragaglia-succi-2005},
or to apply diffuse
scattering~\cite{bib:ansumaili-karlin-2002,bib:sofonea-sekerka-2005,bib:niu-shu-chew-2004}.
It has been shown by Guo et al. that these approaches are virtually
equivalent~\cite{bib:guo-shi-zhao-zheng-2007}. Another possibility is to
modify the fluid viscosity due to local density variations in order to
model slip~\cite{bib:nie-doolen-chen}. In both cases, the parameters
determining the properties at the boundaries are ``artificial'' parameters
and they do not have any obvious physical meaning. Therefore, these parameters are
not easily mappable to experimentally available quantities.  We model the
interaction between hydrophobic channel walls and the fluid by means of a
multiphase lattice Boltzmann model. Our approach overcomes this problem
by applying a mesoscopic force between the walls and the fluid. A similar
approach is used by Zhu et
al.~\cite{bib:zhu-tretheway-petzold-meinhart-2005}, Benzi et
al.~\cite{bib:benzi-etal-06}, and Zhang et al.~\cite{bib:zhang-kwok-04}.
This force applied at the boundary can be linked to the contact angle
which is commonly used by experimentalists to quantitatively describe the
wettability of 
materials~\cite{benzi-etal-06b,bib:huang-thorne-schaap-sukop-2007,bib:jens-schmieschek:2010,bib:jens-kunert-herrmann:2005}.
While we are not aware of further lattice Boltzmann simulations to study the
flow over a bubble mattress, a number of authors has applied
various lattice Boltzmann multiphase and multicomponent models to study
the properties of droplets on chemically patterned and superhydrophobic
surfaces~\cite{kusumaatmaja-06,bib:kusumaatmaja-yeomans-2007,bib:pirat-etal-2008,bib:hyvaluoma-koponen-raiskinmaki-timonen:2007,bib:moradi-varnik-raabe-steinbach-10}.

The lattice Boltzmann method is based on the Boltzmann kinetic equation
\begin{equation}
\label{eq:boltzmann}
\left[\frac{\partial }{\partial t}+{\bf u} \cdot \nabla_{\bf x}\right] f({\bf x,u},t)={\bf \Omega},
\end{equation}
which is discretized on a lattice. The Boltzmann equation describes the evolution of the
single particle probability density $f({\bf x},{\bf u},t)$, where ${\bf x}$
is the position, ${\bf u}$ the velocity, and $t$ the time.
The derivatives on the left-hand side represent propagation of particles in 
phase space whereas the collision operator ${\bf \Omega}$ takes
into account molecular collisions. 
When discretized, to represent the correct 
hydrodynamics, the collision operator should conserve mass and momentum, and it
should ensure sufficient isotropy and be Galilean invariant. 
By performing a Chapman-Enskog analysis, it can be shown that such a
collision operator ${\bf \Omega}$ leads to flow behavior following the Navier-Stokes
equation~\cite{bib:succi-01}. 
In the lattice Boltzmann method the time
$t$, the position ${\bf x}$, and the velocity ${\bf u}$ are discretized
leading to a discretized version of Eq.~\ref{eq:boltzmann}:
\begin{equation}
\label{LBeqs}
\begin{array}{cc}
f_i({\bf x}+{\bf c}_i, t+1) - f_i({\bf x},t) =
\Omega_i, &  i= 0,1,\dots,B.
\end{array}
\end{equation}
Above, $f_i({\bf x},t)$ indicates the amount of fluid 
at site ${\bf x}$ at time step $t$ with velocity ${\bf c}_i$,. Our simulations are
performed on a three dimensional lattice with $B=19$ discrete
velocities (the so-called D3Q19 model).
For the collision operator
$\Omega_i$ we choose the Bhatnagar-Gross-Krook (BGK) form~\cite{bib:bgk}
\begin{equation}
\label{Omega}
 \Omega_i =
 -\frac{1}{\tau}(f_i({\bf x},t) - f_i^{eq}({\bf u}({\bf x},t),\rho({\bf x},t)))\mbox{ ,}
\end{equation}
where $\tau$ is the mean collision time that determines the kinematic viscosity
\begin{equation}
\label{eq:vis}
\nu=\frac{2\tau-1}{6} . 
\end{equation}
of the fluid. In this study the
relaxation time $\tau$ is kept constant at value 1.0. Due to the collisions,
the system relaxes towards an equilibrium distribution
$f_i^{eq}$ which can be derived imposing restrictions on the
microscopic processes, such as explicit mass and momentum conservation.
In our implementation we choose for the equilibrium distribution function
\begin{equation}
\label{eq:equil}
f_i ^{eq} = \zeta_i \rho \left[ 1+\frac{{\bf c}_i \cdot {\bf u}}{c_s^2} 
+ \frac{({\bf c}_i \cdot {\bf u})^2}{2c_s^4} - \frac{u^2}{2c_s^2} \right] ,
\end{equation}
which is a polynomial expansion of the Maxwell distribution where
$\zeta_i$ are the lattice 
weights resulting from the velocity space discretization, and $c_s=1/\sqrt{3}$ is the speed 
of sound for the D3Q19 lattice.
The hydrodynamic quantities are obtained as moments of the single-particle
distribution function $f_i({\bf x},t)$. For example, the density
at lattice site ${\bf x}$ is 
\begin{equation}
\rho({\bf x},t)\equiv \sum_i f_i({\bf x},t),
\end{equation}
and the macroscopic velocity ${\bf u}({\bf x},t)$ is obtained from 
\begin{equation}
\rho({\bf x},t){\bf u}({\bf x},t) \equiv \sum_i
f_i({\bf x},t){\bf c}_i.
\end{equation}

Mean-field interactions between fluid particles are introduced by
following the work of Shan
and Chen, as a mean-field body force between nearest
neighbours~\cite{bib:shan-chen-93,bib:shan-chen-liq-gas},
\begin{equation}
\label{eq:sc}
{\bf F} = G_b \psi({\bf r}) \sum_i \zeta_i
          \psi({\bf r} + {\bf c}_i) {\bf c}_i,
\end{equation}
where $\psi = 1 -\exp{(-\rho/\rho_0)}$ is an effective mass, $G_b$
tunes the strength of the interaction, and $\rho_0$ is a reference density. 
This force term leads to a
non-ideal equation of state with pressure $P = c_s^2\rho +
\frac{1}{2}c_s^2G_b\psi^2$, and it enables simulations of liquid-vapor
systems with surface tension. To model the wetting behavior at fluid-solid
surfaces, a similar interaction is added between the fluid and solid
phases, and the contact angle is tuned by setting a density value $\rho_w$
at the boundaries~\cite{bib:sbragaglia-etal-06,bib:jens-kunert-herrmann:2005}.
Additionally, we apply mid-grid bounce back boundary conditions between
the fluid and the surface which assures vanishing velocities at solid
surfaces. Here, a distribution function that would be advected into a
solid node is simply reversed and advected into the opposite
direction~\cite{bib:succi-01}.

From molecular dynamics simulations it is known that the fluid-wall
interactions causing a slip phenomenon usually take place within a few
molecular layers of the liquid along the boundary
surface~\cite{bib:thompson-troian-1997,bib:cieplak-koplik-banavar-01,bib:cottin-bizone-etall-04,baudry-charlaix-01}.
Our coarse-grained fluid wall interaction acts on the length scale of one
lattice constant and does not take the molecular details into account.
Therefore, coarse-grained implementations based on the lattice Boltzmann
method are only able to reproduce an averaged effect of the interaction
and cannot fully resolve the correct flow profile very close to the wall
and below the resolution of a single lattice spacing. However, in order
to understand the influence of the hydrophobicity on experimentally
observed apparent slip, it is sufficient to investigate the flow behavior
on more macroscopic scales as they are accessible for experimental
investigation. Coarse-grained interaction models could be improved by a
direct mapping of data obtained from MD simulations to the coupling
constant $G_b$ allowing a direct comparison of the influence of
liquid-wall interactions on the detected
slip~\cite{bib:jens-kunert-herrmann:2005}. Similar approaches are known
from quantitative comparisons of lattice Boltzmann and molecular dynamics
simulations in the
literature~\cite{bib:horbach-suci-2006,bib:chibbaro-etal-2008}.

In recent years we studied the influence of a number of parameters on an
apparent boundary slip using the method described above. Here, we shortly
review the main conclusions of our previous findings. A more comprehensive
review can be found in~\cite{bib:jens-kunert-jari:2010}.

In~\cite{bib:jens-kunert-herrmann:2005}, we showed that
our mesoscopic approach is able to reach the small flow velocities of
known experiments and reproduces results from experiments and other
computer simulations, namely an increase of the slip with increasing
liquid-solid interactions, the slip being independent of the flow
velocity, and a decreasing slip with increasing bulk pressure~\cite{bib:jens-kunert-herrmann:2005}.

If typical length scales of the system are comparable to the
scale of surface roughness, the effect of roughness cannot be neglected
anymore.  The influence of surface variations on the slip length $b$ has
been investigated by numerous authors. It was de\-mon\-stra\-ted by
Richardson that roughness leads to higher drag forces and thus to no-slip
on macroscopic scales~\cite{bib:richardson-73}. An experimental confirmation was later
presented by McHale and Newton~\cite{mchale-newton-04}. Sbragaglia et al.
applied the LB method to simulate fluids in the vicinity of
microstructured hydrophobic surfaces~\cite{bib:sbragaglia-etal-06},
Al-Zoubi et al. demonstrated that the LB method is well applicable to
reproduce known flow patterns in sinusoidal
channels~\cite{bib:alzoubi-brenner-2008}, and Varnik et
al.~\cite{varnik-raabe-06,varnik-dorner-raabe-06} have shown that even in
small geometries rough channel surfaces can cause flow to become
turbulent.
Recently, we presented the idea of an effective wall position at which the
no-slip boundary condition holds for rough channel
surfaces~\cite{bib:jens-kunert:2007b}. We investigated the influence of
different types of roughness on the position of the effective boundary
$h_{\rm eff}$.  Further, we have shown how the effective boundary depends
on the distribution of the roughness elements and how roughness and
hydrophobicity interact with each other~\cite{bib:jens-kunert:2008c}.  We
were also able to simulate flow over surfaces generated from AFM data of
gold coated glass used in microflow experiments by Vinogradova and
Yakubov~\cite{vinogradova-yakubov-06}. We found that the height
distribution of such a surface is approximately Gaussian and that a randomly arranged
surface with a similar distribution gives the same result for the position
of the effective boundary although in this case the heights are not
correlated~\cite{bib:jens-kunert:2007b}. Currently, we model AFM based
measurements to determine lubrication forces in order to
probe the fundamental concept of boundary conditions leading to a slippage or
a shift of the effective boundary
position~\cite{bib:jens-kunert-vinogradova:2010,bib:jens-kunert-2009}.

A natural continuation of our previous works on roughness induced apparent
boundary slip is the analysis of flow along superhydrophobic surfaces as
presented in this article.  It has been recently
predicted~\cite{bib:cottin-bizone-etall-03} and experimentally
reported~\cite{bib:perot-rothstein-04} that the so-called Fakir effect or
Cassie state considerably amplifies boundary slippage on superhydrophobic
surfaces. Using highly rough hydrophobic surfaces such a situation can be
achieved. Instead of entering the area between the rough surface
elements, the liquid remains at the top of the roughness and traps air in
the interstices. Thus, a very small liquid-solid contact area is
generated. In this article we quantify the slip in such systems using
Couette and Poiseuille flow, where the flow is confined between two
parallel walls separated by a distance $2d$. In our simulations, the lower
wall is static and is patterned with holes or grooves to which vapor
bubbles are trapped.  The upper one is smooth and can be driven with shear
velocity $u_0$ in the case of a Couette setup. The system boundaries are
periodic and a unit cell of the regular array is included in a simulation.
To trap bubbles to holes, some heterogeneity is needed at the edges of
holes in order to pin the contact line. To this end, we use different
wettabilities for boundaries in contact with the main channel and with the
hole.  For a Couette setup, the effective slip length can be calculated
from the shear stress $\sigma = \mu {\rm d}u/{\rm d}z$ acting on the upper
wall, which is obtained from the no-slip boundary condition imposed at the
fluid-solid boundaries. Thus, the effective slip length reads as $b = \mu
u_0/ \sigma - 2d$, where $\mu$ is the dynamic viscosity of the liquid.  In
case of a Poiseuille flow, a driving body force is applied effectively
generating a pressure gradient $\frac{\partial P}{\partial z}$.  Assuming
Navier's boundary condition, $b$ is measured by fitting the theoretical
velocity profile as given by
\begin{equation}
\label{eq:profil03}
u_z(x)=\frac{d^2 }{2 \mu} \frac{\partial P}{\partial z}\left[\frac{
-x^2}{2d^2}+ \frac{x}{d+b}+\frac{2b}{d+b} \right]
\end{equation}
to the simulated data via the slip length $b$.
This flow profile assumes the lower boundary to have a slip while on the
upper wall a no-slip boundary condition is applied.

\begin{figure}[htb]
\centerline{\includegraphics[width = 0.25\textwidth]{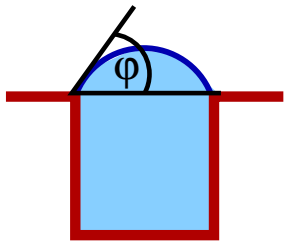}}
\caption{\label{fig:prot}
Definition of the protrusion angle $\varphi$.
}
\end{figure}

The parameter $G_b$ is chosen such that
the density ratio between liquid and gas is between 22 and 44.
This ratio is too small for a realistic description of gas bubbles in a
liquid. Also, the interface between both phases is of finite width causing
the resistance in the vapor phase.  However, these limitations of
multiphase LB models do not influence the qualitative insight obtained
from the simulations.

In equivalence to the definition of the contact angle, which is not
suitable for bubble matresses as studied in this paper, we utilize the
so-called protrusion angle $\varphi$ to classify the bubbles in the simulated systems.
The protrusion angle is defined by measuring the angle between the surface
and a tangent at the bubble surface (see Fig.~\ref{fig:prot}). The
protrusion angle can be varied in the simulations by changing the bulk
pressure of the liquid phase.
\section{Results}
\subsection{Surfaces patterned with microgrooves}
\begin{figure}[hbt]
\centerline{\includegraphics[height = 0.4\textwidth]{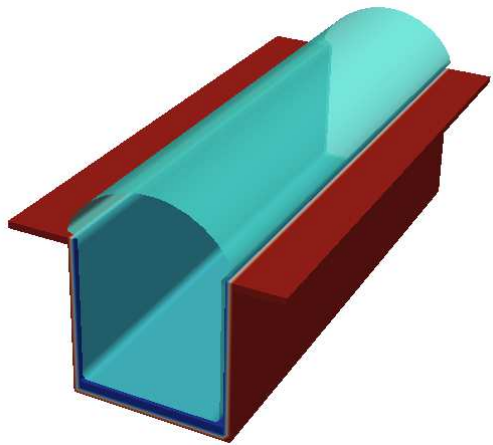}}
\caption{\label{fig:3Dgroove}
3D visualization of a grooved surface. The width of the unit cell is $L$
and the width of a groove is defined by $c$.
}
\end{figure}

A surface structure that has been studied both experimentally and theoretically
in great detail consists of longitudinal grooves etched in a smooth
surface. If such grooves are hydrophobic, they can be filled with vapor
forming cylindrical bubbles that are surrounded by liquid. The surface
between individual grooves is hydrophilic causing an effective pinning of
the three-phase contact line. In simulations, such hydrophilic surfaces can be obtained by
setting a virtual surface density of the same order as the liquid density
when computing the fluid-surface interactions as given by Eq.~\ref{eq:sc}.
The hydrophobic interactions are generated in a similar fashion by
choosing a substantially smaller value for the virtual surface density.
Fig.~\ref{fig:3Dgroove} shows
a 3D visualization of the simulated unit cell. The length of the system
is $L=40$ lattice units and the width of the groove is defined by $c$
resulting in a distance between two grooves given by $(L-c)$ lattice units.
The height of the channel is $d=135$ lattice units in $x$-direction, where
the top of the system is defined by a smooth no-slip surface. The flow is
generated by a constant body force acting in $z$-direction effectively
resulting in a Poiseuille profile. Apart from the top and bottom plane,
all boundaries are periodic. 

A similar system was investigated in the experiments by Tsai et
al.~\cite{tsai-etal-09}. They studied the flow field over gas filled grooves
oriented in flow direction by using micro particle image velocimetry
(micro PIV). In~\cite{tsai-etal-09} they presented measured flow profiles
and the corresponding effective slip lengths. The behavior of the slip
length in dependence on different geometrical parameters has been
investigated and it was concluded that the obtained values for slip length cannot
be explained by the theory of Philip~\cite{philip-72} which provides an
analytical model for slip flow on striped full-slip / no-slip surfaces.
Their explanation was that due to the meniscus forming the bubble surface
the no-shear (i.e. full-slip) parts develop an additional drag, similar to
the effect reported by Richardson who showed that a rough no-shear surface
creates sufficient drag to obtain macroscopically a no-slip
boundary~\cite{bib:richardson-73}.

\begin{figure}[hbt]
\centerline{\includegraphics[width = 0.35\textwidth,angle=270]{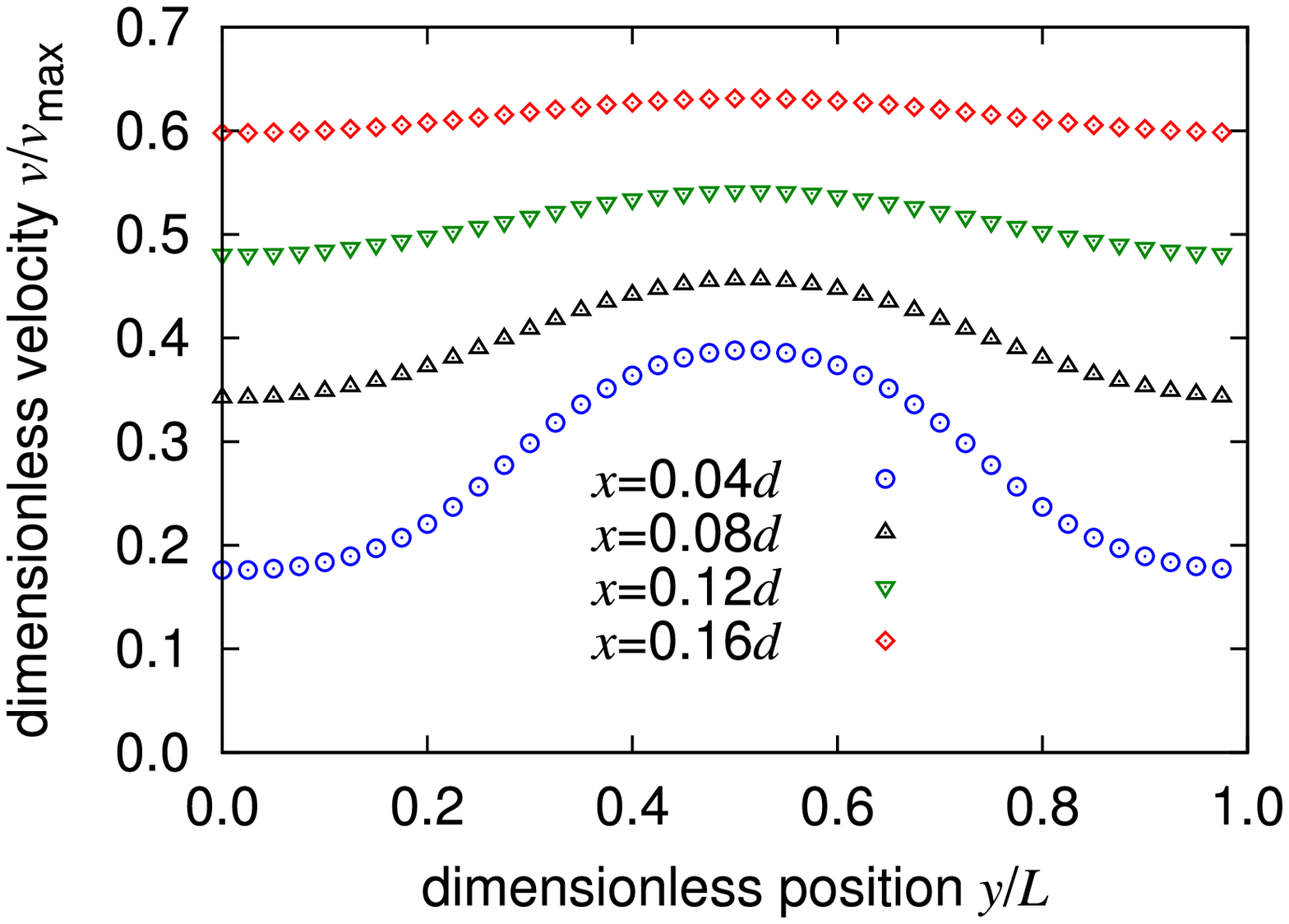}
\includegraphics[width = 0.35\textwidth,angle=270]{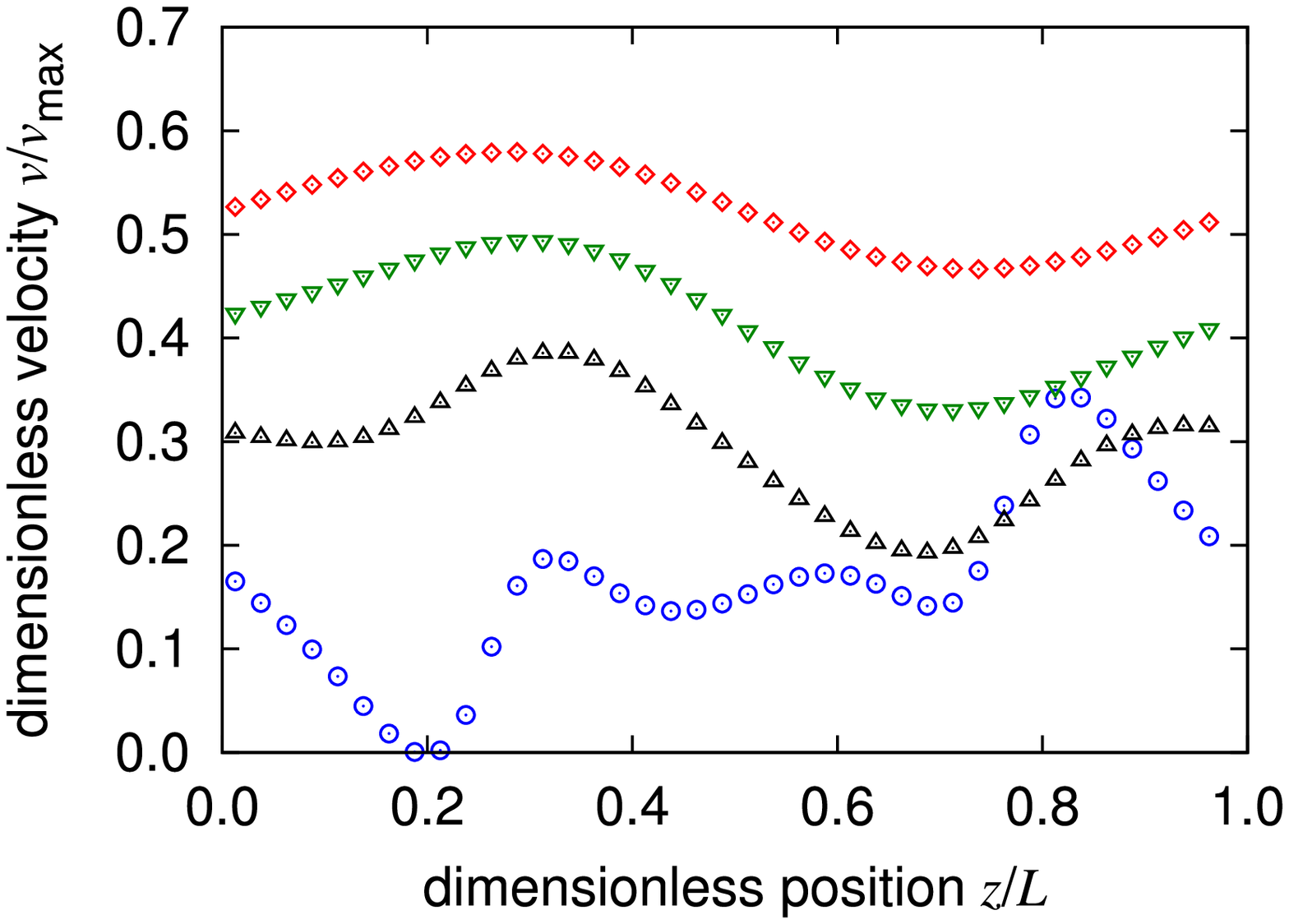}}
\caption{\label{fig:profile}
Flow profile perpendicular to the grooves for flow parallel (left) and
perpendicular (right) to the grooves at different $x$ positions over the
surface. The groove is located in the centre of the system between $0.25L$
and $0.75L$. In the left panel a strong increase of the
velocity in close vicinity to the bubble can be seen that relaxes within $0.16d$ to
an almost undisturbed flow. The right plot demonstrates that flow perpendicular
to the grooves causes more significant disturbances, which get still damped
rather quickly.
}
\end{figure}

Following the work of Tsai et al. we first present simulations of flow
over grooves oriented in flow direction. The width of the groove is chosen
to be $c/L=0.5$. The protrusion angle in the experiments and in the
simulations is chosen to be $\varphi=-10^{\circ}$, i.e., the liquid-gas
interface forms a dimple rather than a bubble. The meniscus increases the
vapor covered area (no-shear surface) but also introduces roughness that
leads to a higher friction and therefore a reduced slip. In
Fig.~\ref{fig:profile} we show the flow profile over one groove for
different distances from the surface.  At a very close distance $x=0.04d$
the bubble has a strong influence on the flow, since the velocity in the
centre of the groove is more than twice as large as the velocity in the
centre of the void space. This effect is damped very quickly, i.e., at
$x=0.16d$ the velocity increase in the area above the bubble is less than
$10\%$ compared to the undisturbed case. The strong damping allows the
treatment of such a grooved surface as a surface with an effective
boundary position.  The flow profiles are consistent with the results of
Tsai et al. who also found a strong increase of the velocity above the grooves that is
damped strongly further in the bulk.  Further, the measured and the
simulated slip lengths $b/L$ are of the order of half the theoretical
prediction by Philip~\cite{philip-72}. The reason for this is the
assumption of a flat surface with stripes of no-slip and no-shear, while
in the case of vapor filled grooves a meniscus is formed introducing an
additional roughness that reduces the slip of the fluid over the surface. 

In order to compare our results to the theoretical results of Lauga and
Davis~\cite{bib:davis-lauga-2009}, we also consider grooves which are
oriented perpendicular to the flow.  In the right panel of
Fig.~\ref{fig:profile}, we show the flow profile in the vicinity of a
perpendicular groove as it is assumed by the theory.  Interestingly, here
the flow velocity does not increase as strongly as in the aligned case
above the groove but right after it. Again, a deceleration in front of the
groove can be observed and the flow profiles are not symmetric towards the
centre of a unit cell which is caused by the driving direction of the
flow. The disturbances caused by the bubble are not damped as quickly as
in the case of longitudinal grooves. For example, even at $x=0.16d$ the
flow velocity changes between $0.45v_{\rm max}$ and $0.59 v_{\rm max}$. 
Further the disturbance close to the boundary is much stronger. At $x=0.04d$
one can see a strong decrease of the velocity at the beginning of the groove
and a strong increase at the end.
The reason is the dimple shape of the meniscus. This leads to a compression of the
streamlines at the end of the meniscus which is equivalent to an increased velocity.
The dependence of the slip length on the orientation is consistent with the work of Bazant et al. who have shown for flat stripes that a generalized tensor form would
be required to describe the surface properly~\cite{Bazant08}.

The shape of the meniscus has a significant influence on the slip
length. The above mentioned analytical approach of Davis and Lauga
described the 
effective slip length in dependence on the protrusion angle $\varphi$ on a
surface with grooves perpendicular to the flow
direction~\cite{bib:davis-lauga-2009}. The theory assumes rigid bubbles
with a full-slip surface, which corresponds an infinitely thin liquid-gas
interface and vanishing gas density. In this case the slip length is given by
\begin{equation}
\frac{b}{c}=\pi\left(\frac{c}{L}\right) \int_0^\infty A(s) {\rm d}s,
\label{eq:lauga}
\end{equation}
with
\begin{equation}
A(s)=\frac{s}{\sinh 2s(\pi-\varphi) + s \sin 2 \varphi}\times \left[\cos 2
  \varphi +   \frac{s \sin 2 \varphi \cosh s \pi + \sinh s(\pi - 2\varphi)}{\sinh s \pi}  \right].
\end{equation}
The only parameters entering the calculation of the effective slip length
$b$ are the protrusion angle $\varphi$ and the ratio between the width of
a groove $c$ and the length of a unit cell $L$. The actual values of the
slip length have to be calculated numerically. 

\begin{figure}[htb]
\centerline{\includegraphics[width=0.4\textwidth,angle=270]{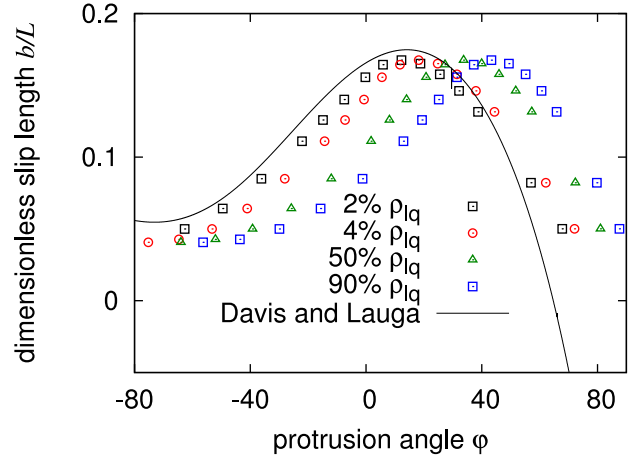}}
\caption{\label{fig:protangle} Slip length versus protrusion angle
$\varphi$ for different threshold values. The groove has a width of $c=30$
and the system length is $L=40$. Since the multiphase lattice Boltzmann
model used is a diffuse interface model, the protrusion angle is not
defined uniquely but depends on where the actual interface position is
assumed, i.e. which threshold $\rho_t$ is chosen.  Due to the non perfect
slip on the vapor phase the analytical solution by Lauga et al. is scaled
down by a factor of $0.75$.
}
\end{figure}
In Fig.~\ref{fig:protangle} we show the dimensionless slip length $b/L$
versus the protrusion angle $\varphi$ as given by Eq.~\ref{eq:lauga} and
by the simulation. The system length in this case is $L=40$ lattice units
and the width of a groove is $c=30$ lattice units. The protrusion angle is
measured by fitting a circle to the cross section of the interface.  Since
the interface between the vapor and the liquid is diffuse, the actual
bubble position is not strictly defined. This is in contrast to the
theoretical solution and renders the determination of the protrusion angle
difficult.  Therefore, we apply different threshold values $\rho_t$ for
the fluid density at the interface to determine the protrusion angle. This
value has to be somewhere between the high (liquid) density
$\rho_{lq}=2.2$ and the lower (gas) density $\rho_{g}=0.05$. In
Fig.~\ref{fig:protangle} the effect of choosing different threshold values
between $\rho_t=0.02\rho_{\rm lq}$ and $0.9\rho_{\rm lq}$ is demonstrated.
Here, it can be observed that a variation of the threshold can lead to a
shift in the protrusion angle of more than $20^{\circ}$. 

By comparing the simulation data to the results of Davis and Lauga
Eq.~\ref{eq:lauga}, we find that the qualitative shape of the curve is
well reproduced, i.e., the slip length first increases with rising
$\varphi$ up to a maximum at $\varphi_{\rm max}$ and then follows a steep
decrease for high $\varphi$. As will be shown later, it eventually can
even become negative.  However, in addition to the possible variation of
$\varphi$ we find a second deviation between theory and simulation: the
detected slip length $b$ is lower than predicted by Davis and Lauga.  To
be able to compare theory and simulation, the theoretical values are
scaled by a factor of $0.75$ to fit the data.  This can be explained by
the fact that the diffuse interface must not be described by a smooth full
slip cap. Instead it shows a finite slip due to the friction within the
interface region. Further, the density ratio between liquid and vapor is
limited in the lattice Boltzmann model used. In the presented case it is
only $1/44$. Therefore, the shear resistance on the bubble surface is only
reduced by this factor, while in a real system consisting of, e.g.,  water
and air, this ratio would be of the order of $1/1000$ rendering the
assumption of no shear more realistic. Apart from the shift due to the
finite interface width, the simulation is able to recover the main
conclusions from the theory. 

\begin{figure}[htb]
\centerline{\includegraphics[width=0.4\textwidth,angle=270]{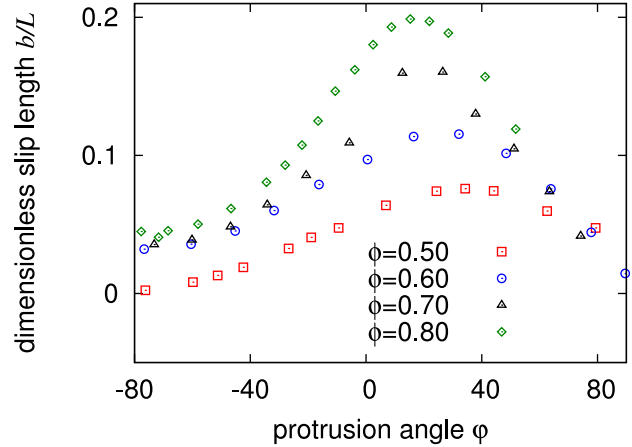}}
\caption{\label{fig:laugawide}
Normalized slip length $b/L$ versus the protrusion angle $\varphi$, for different
ratios between the width of the grooves $c$ and the system length $L$. 
The behavior is similar to the prediction of Eq.~\ref{eq:lauga}. Further, a larger
$\phi=c/L$ leads to a strong increase of the slip length. 
}
\end{figure}

We further investigate the influence of the area covered by the bubble
$\phi=c/L$ on the slip length. Different channel widths are been
investigated for different protrusion angles $\varphi$. The threshold
value determining the interface position was chosen to be
$\rho_t=0.04\rho_{lq}$. Our results are shown in Fig.~\ref{fig:laugawide}.
The qualitative behavior of the results again follows Eq.~\ref{eq:lauga}.
The maximum slip length is increased very strongly by increasing the
relative width of the groove.  An increase of the surface coverage from
$\phi=0.5$ to $0.8$ leads to an increase of the dimensionless slip length
where the values for the respective maxima range from $b/L=0.07$ to
$0.19$. Such a strong increase is also predicted by Eq.~\ref{eq:lauga}.
Further, due to the influence of the increased roughness for
$\varphi>60^{\circ}$  the slip length becomes smaller for larger
$\varphi$. Also for $\varphi<-40^{\circ}$ the slip becomes smaller than
$b<0.05L$ and nearly independent on the surface coverage $\phi$. 

To conclude, in the current section we have introduced our simulation
methodology and demonstrated that it is well suited to reproduce
experimental flow profiles and to study the influence of the protrusion
angle $\varphi$ on a detected effective slip. However, it has to be taken
into account that the simulation model is a diffuse interface method
effectively rendering our bubbles to be on the nanoscale. A comparison with the
theoretical description of Davis and Lauga demonstrates that the diffuse
interface can have a strong effect on the actual slip which has to be
considered when designing surfaces with nanoscale patterns for specific
applications.

\subsection{Spherical nanobubbles}
\label{prlreview}
Steinberger et al. utilized surfaces patterned with a square array of
cylindrical holes to demonstrate that gas bubbles present in the holes may
cause a reduced
slip~\cite{bib:steinberger-cottin-bizonne-kleimann-charlaix:2007}.
Numerically, they found even negative slip lengths for flow over such
bubble mattresses, i.e., the effective no-slip plane is inside the channel
and the bubbles increase the flow resistance. Motivated by Steinberger's
work, we consider negative slip lengths on bubble covered surfaces and
also discuss the question of shear-rate dependent slip. In particular, we
demonstrate that bubbles can generate a shear-rate dependence of the
detected slip length.  The current section is a review of the studies
presented in~\cite{bib:jens-jari:2008,bib:jens-kunert-jari:2010} and here
we relate them to recent references in the literature and newer findings
as presented in remainder of this contribution.

The distance between top and bottom walls is $d=40$ lattice nodes in all
simulations in this section and
the area fraction of holes is kept at $0.43$. Similar to the previous section, a
unit cell of the regular array is included in a simulation and periodic
boundary conditions are applied at domain boundaries. 

\begin{figure}[htb]
\centerline{\includegraphics[height=0.3\textwidth]{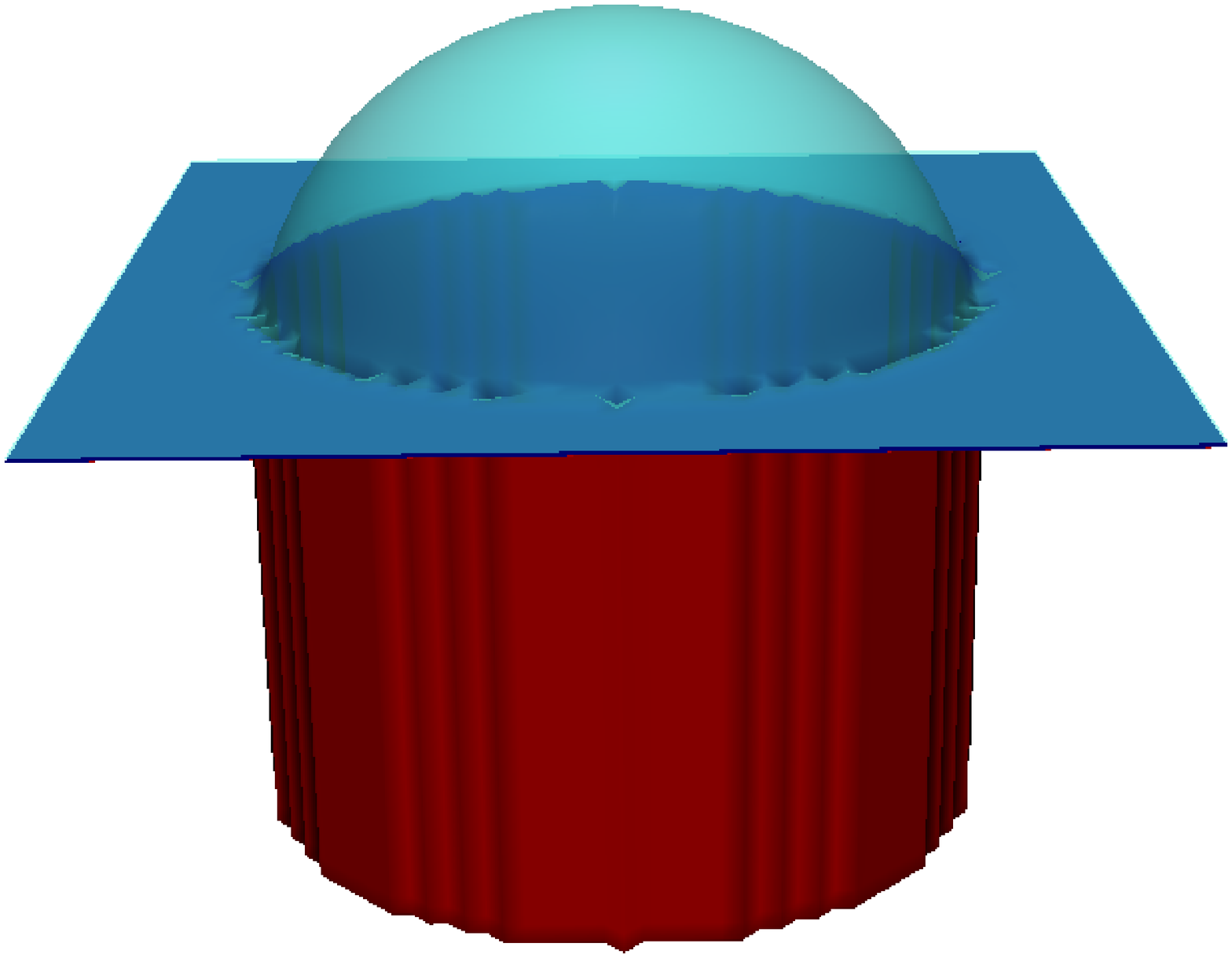}
\hspace{0.0 \textwidth}
\includegraphics[height=0.3\textwidth]{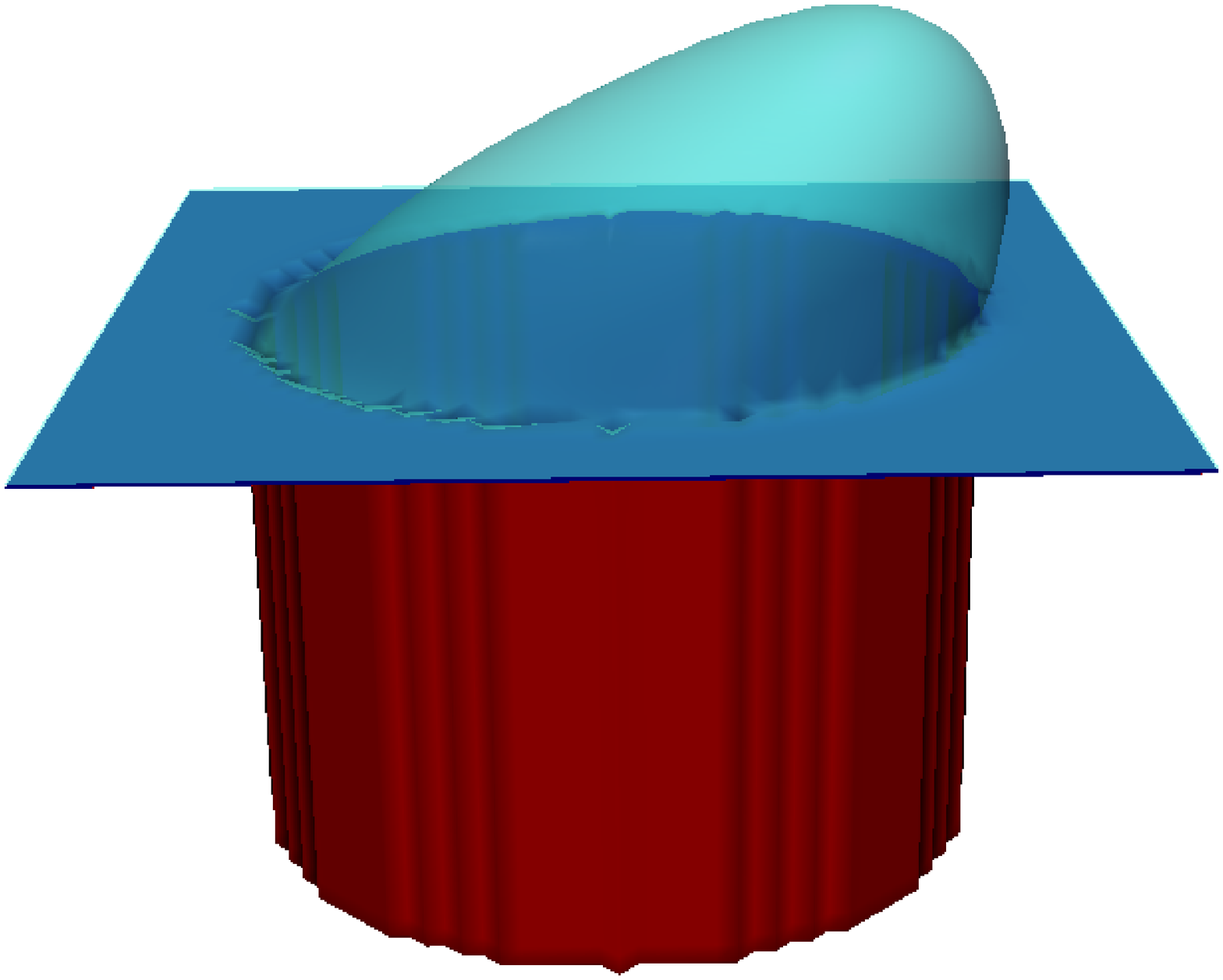}}
\caption{\label{fig:3dsphere}
3D visualization of a spherical bubble. While the left panel shows the
undisturbed shape, the right panel demonstrates the deformation of the
bubble under strong shear.}
\end{figure}
\begin{figure}[htb]
\centerline{\includegraphics[width=0.4\textwidth,angle=270]{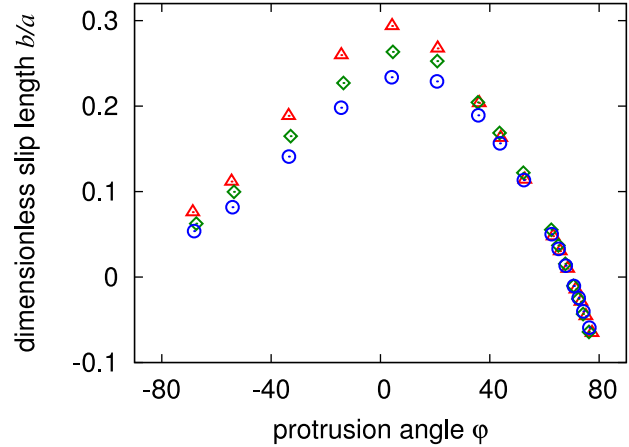}}
\caption{\label{fig:sphereprot}
Slip length $b$ as a function of the protrusion
angle $\phi$. 
Corresponding results are given for a rhombic array (triangles),
a rectangular array (diamonds), and a square array (circles).}
\end{figure}

We investigate the effect of a modified protrusion angle and different
surface patterns by using square, rectangular, and rhombic bubble arrays.
The cylindrical holes have a radius $a = 20$ lattice units and the area
fraction of the holes is equal in all cases. The shear rate is such that
the Capillary number $Ca = \mu a G_s /\gamma = 0.16$. Here, $G_s$ and
$\gamma$ are the shear rate and surface tension, respectively. A snapshot
of a simulated system is shown in the left part of
Fig.~\ref{fig:3dsphere}.  The slip lengths obtained for differently shaped
unit cells and protrusion angles are shown in Fig.~\ref{fig:sphereprot}.
Following the discussion on the diffuse interface and the definition of
the protrusion angle as given in the previous section, we choose the
threshold value for the density defining the interface to be at
$\rho_t=0.5\rho_{\rm lq}$.
The observed behavior is similar to that reported
in~\cite{bib:steinberger-cottin-bizonne-kleimann-charlaix:2007}, where a
square array of holes was studied. In particular, we observe that when
$\varphi$ is large enough $b$ becomes negative. Moreover, for the case
presented in the current paper the slip length is maximized when the
protrusion angle is close to zero. In practical applications, this would allow
the highest possible throughput in a microchannel to be obtained. The
behavior of the slip length can be explained by thinking of an increased
surface roughness if the protrusion angle is larger or smaller than zero
which is in competition with the increased area of low friction in the
case of large bubbles. Since the area fraction of the bubbles is kept
constant for all three different unit cells, our results clearly indicate
that slip properties of the surface can be tailored not only by changing
the protrusion angle but also by the array geometry. In the presented
study, the highest slip lengths are obtained for the rhombic unit cell. 

Generally, our findings are in accordance to the results presented in the
previous section as it has been confirmed in the theoretical paper of
Davis and Lauga~\cite{bib:davis-lauga-2009}. Obviously, also here the
limitations of the comparison due to the diffuse interface and finite
density ratio of the liquid and gas phases still hold. In addition,
the simulated system is now three dimensional while Davis and Lauga
only considered a two-dimensional case.

\begin{figure}[htb]
\centerline{\includegraphics[width=0.4\textwidth, angle= 270]{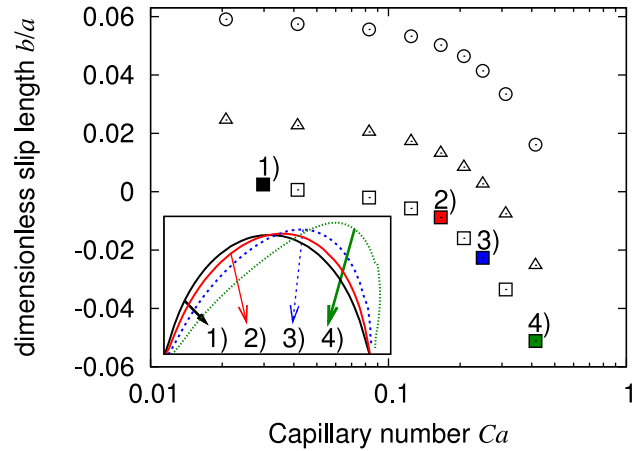}}
\caption{\label{fig:vis}
The slip length $b$ as a function of the capillary
number $Ca$ for a square array of spherical bubbles with three different
protrusion angles, $\varphi = 63^\circ, 68^\circ,$ and
$71^\circ$ (from uppermost to lowermost) is shown.
The inset depicts cross sections of liquid-gas interfaces
for four capillary numbers and $\varphi=71^\circ$~\cite{bib:jens-jari:2008}.
}
\end{figure}
In the following, the shear-rate dependence of the slip length is
investigated. As the shear rate and thus the viscous stresses increase the
bubbles are deformed and the flow field is modified. An example of such a
deformed bubble is shown in Fig.~\ref{fig:3dsphere} (right).  In
Fig.~\ref{fig:vis}, we show the simulated normalized slip length $b/a$ as a function of
the Capillary number for three different protrusion angles.  The Capillary
numbers chosen are at the higher end of the experimentally available range.
The inset of Fig.~\ref{fig:vis} shows cross sections of the liquid gas
interface of a bubble with protrusion angle $\varphi=71^\circ$
demonstrating the correlation between shear rate and deformation.  Our
lattice Boltzmann simulations clearly show shear-rate dependent slip, but
the behavior is opposite to that found in some experiments: in fact, the
slip lengths measured by us decrease with increasing shear due to a
deformation of the bubbles which effectively reduces the roughness of the
surface. In the experiments, surface force apparatuses are used (see,
e.g., Ref.~\cite{bib:zhu-granick-01}), where a strong increase in the slip
is observed after some critical shear rate. This shear-rate dependence has
been explained, e.g., with formation and growth of
bubbles~\cite{bib:degennes-02,lauga-brenner-04}. In our simulations, there
is no formation or growth of the bubbles as we only simulate a steady case
for given bubbles. The experiments on the other hand are dynamic. However,
our results indicate that the changes in the flow field which occur due to
the deformation of the bubbles cannot be an explanation for the shear-rate
dependence observed in some experiments. Our results are consistent
with~\cite{bib:jens-kunert:2007b}, where it is shown that smaller
roughness leads to smaller values of a detected slip.  In the present
case, the shear reduces the average height of the bubbles and thus the
average scale of the roughness decreases as well.  However, a possible
explanation of the dependence of the slip length on the shear rate was
recently given by Gao and Feng~\cite{bib:gao-feng:2009}. They argue that
the experimentally observed increase of $b$ can be explained by a
depinning of the three-phase contact line resulting in the extreme case in
a continuous gas film on the surface. Such a behavior is omitted in our
simulations since the interface between liquid and gas favours to stay at
the edge of a hole due to the different wettabilities inside the holes and
on the top of the surface which effectively results in a contact-line
pinning.

\subsection{Bubble-size distributions}
As already stated in the introduction, in all previous contributions to
the literature concerning slip flow over a bubble mattress, only
mattresses which consist of evenly sized bubbles are
taken into account. However, such a situation is not necessarily realistic
since in reality some kind of imperfections are inevitably present.
Therefore, this section focuses on mattresses with bubbles of different
size. For the sake of simplicity we restrict ourselves to two different
sizes only.  In practice, such a situation can be achieved by structuring
the surface with holes that have different radii. The height of the 
bubbles occurring on such
a structuring can be tuned by variation of the bulk pressure. For a given
bulk pressure, bubbles located in larger holes protrude stronger to the
flow channel than those in the smaller ones. We show that the bubble-size
distribution can significantly affect the slip behavior and, e.g.,
decrease the negative slip caused by strongly protruding bubbles. While
this reduction may hamper experimental observations of negative slip, it
is an attractive candidate for designing bubble surfaces aiming to maximal
flow slippage.

\begin{figure}[htb]
\centerline{\includegraphics[height = 0.4\textwidth]{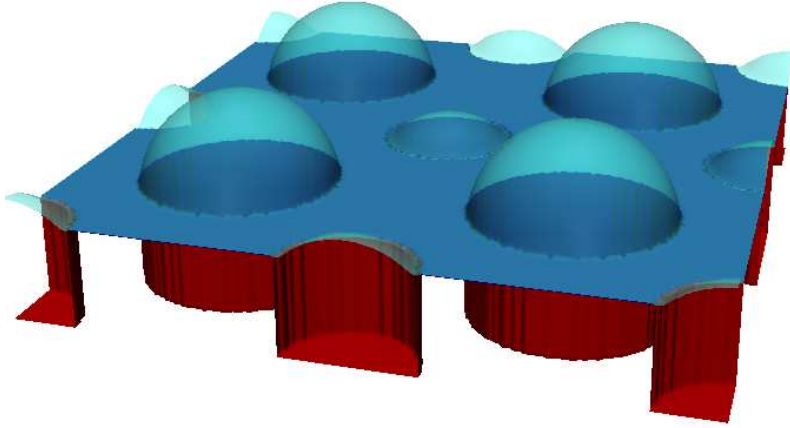}}
\caption{\label{fig:bubbles} 
Visualization of the simulation setup. Four unit cells are shown. Each has
small bubbles (radius $a_2$) at four edges of the cell connected through
periodic boundary conditions, and a large bubble at the centre of the cell
(radius $a_1$). Thus, the bubble mattress consists of two nested square
arrays of bubbles. In the case shown $a_2/a_1 = 0.67$. The upper wall is
smooth and it is moved with constant speed to achieve a shear flow.
}
\end{figure}
The geometrical setup used in the present simulations is shown in
Fig.~\ref{fig:bubbles}. As in the previous section we simulate a Couette
flow between two parallel plates, where one of the plates is smooth and
another is patterned with an array of cylindrical holes. The holes form
two nested square arrays such that one of the arrays consists of holes
with radius $a_1$ and the other one of holes with radius $a_2$. A set of
four unit
cells of the system is shown in Fig.~\ref{fig:bubbles}. In all simulations,
$a_1$ is fixed to $30$ lattice units and $a_2$ varies between $20$ and
$30$ lattice units. The system size is adjusted accordingly
such that the area fraction of the holes is $\phi=0.42$ regardless of the
value of $a_2$. Notice that when the radii $a_1$ and $a_2$ are unequal,
the protrusion angles related to the respective holes are also unequal for
a given value of the bulk pressure. 

\begin{figure}[htb]
\centerline{\includegraphics[width = 0.4\textwidth,angle=270]{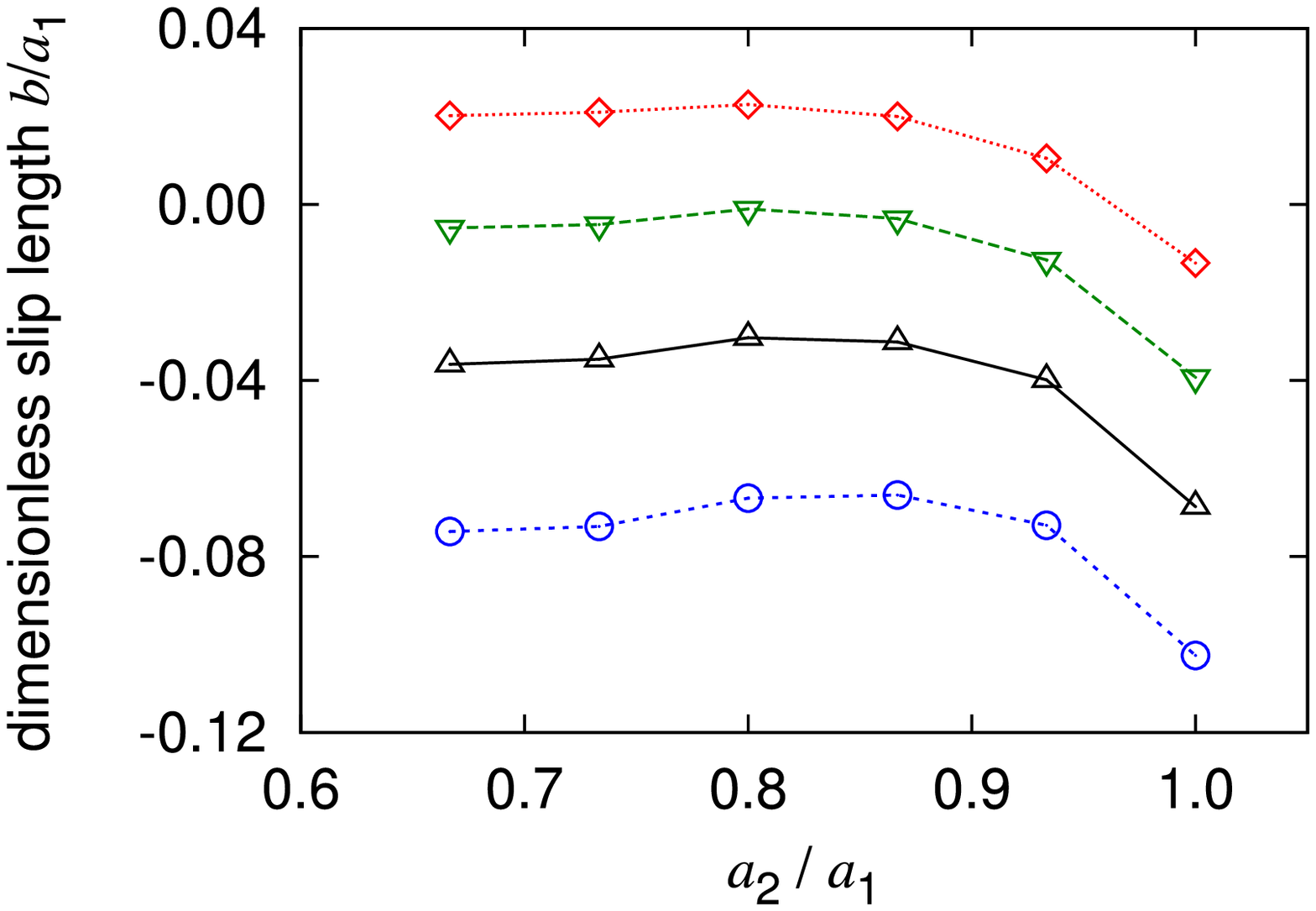}}
\caption{\label{fig:fixCa} 
Slip length as a function of the ratio $a_2/a_1$. The curves are for
different values of bulk pressure such that the protrusion angle of the
larger bubbles are $\varphi = 80^\circ$, $76^\circ$, $73^\circ$, and
$70^\circ$ from bottom to top. The shear rate is fixed and corresponds to
capillary number $Ca = 0.11$.}
\end{figure}
\begin{figure}[htb]
\centerline{\includegraphics[width = 0.4\textwidth,angle=270]{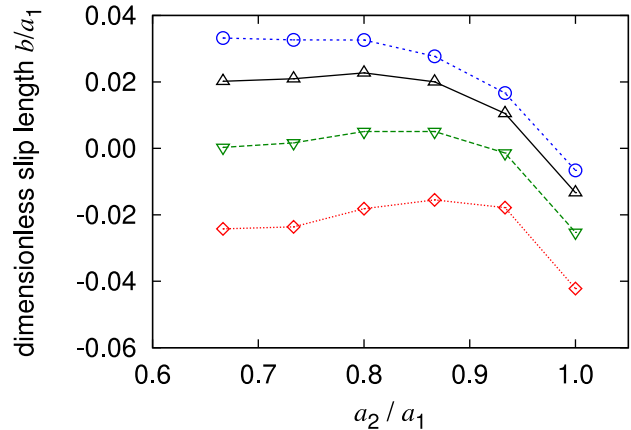}}
\caption{\label{fig:fixPhi} 
Slip length as a function of the ratio $a_2/a_1$. The curves are for different
shear rates corresponding to $Ca$ = $0.036$, $0.11$, $0.18$, and $0.25$
from top to bottom. The bulk pressure is fixed such that the protrusion angle of
the large bubbles is $\varphi = 70^\circ$.
}
\end{figure}

First we investigate how a variation of the size of the smaller holes
affects the effective slip. As the radius $a_2$ is reduced, also the
protrusion angle of the bubbles attached to the smaller holes decreases, i.e.,
these bubbles become flatter. In previous investigations
\cite{bib:steinberger-cottin-bizonne-kleimann-charlaix:2007,bib:jens-jari:2008,bib:davis-lauga-2009}
it has been shown for the case of
mono-sized protruding bubbles, that when the protrusion angle decreases,
the slip
length is increased due to the reduced roughness of the bubble mattress. On
the basis of these findings, one could expect that when the radius $a_2$ is
reduced, the slip length decreases despite the fact that the radius of the
larger holes is kept constant. The simulation results, as shown in
Figs.~\ref{fig:fixCa} and \ref{fig:fixPhi}, show that this is indeed the case 
for small changes in the size of smaller holes. However, when the size
difference between larger and smaller holes is increased such that
$a_2/a_1 \approx 0.8$, a maximum in the effective slip length is observed and
thereafter, for even smaller values of $a_2/a_1$, the slip length starts to
decrease. This observation can be understood as a competition of two different
factors affecting the slip phenomenon. The first one is related
to the above described flattening of the smaller bubbles, which dominates at
ratio $a_2/a_1$ close to unity. However, as the size of the smaller holes is
further reduced, the area fraction of the smaller bubbles also decreases
as well as their effect on the total effective slip. On the same basis, one
might expect that as $a_2/a_1 \rightarrow 0$, slip approaches the same value 
as for  $a_2/a_1=1$. However, as the former case corresponds to a square array of
bubbles and the latter to a rhombic one, the actual values of slip length may be
different as shown in Ref.~\cite{bib:jens-jari:2008} or
Sec.~\ref{prlreview}. 

\begin{figure}[htb]
\centerline{\includegraphics[width = 0.4\textwidth,angle=270]{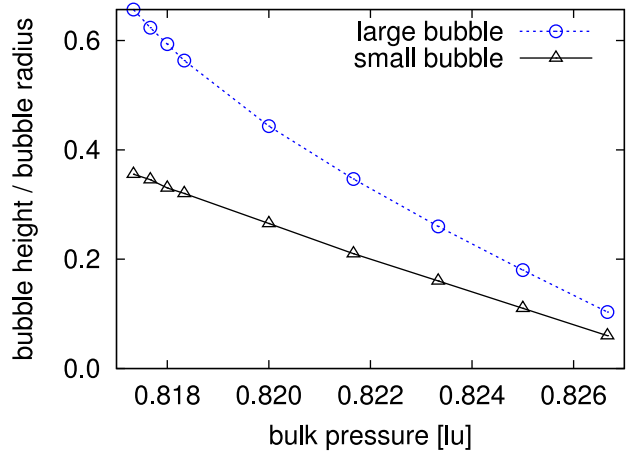}}
\caption{\label{fig:height} 
Bubble height in large and small holes with different bulk pressures
(given in lattice units).
}
\end{figure}

In Fig.~\ref{fig:fixCa} we show how the bulk pressure affects the slip length. 
One can see that when the pressure is increased (i.e., the protrusion
angle $\varphi$ is decreased), also the effective slip length increases. This is due to
the decreasing overall roughness as the height of all bubbles in the system is
reduced. From Fig.~\ref{fig:fixCa} we also observe that the maximum of the
slip length becomes less pronounced and shifts towards smaller values of ratio
$a_2/a_1$ as the bulk pressure increases. This behavior is reasonable as in
this case the difference in the heights of larger and smaller bubbles is
reduced (see Fig.~\ref{fig:height}). 

Finally we consider the effect of shear rate on the slip length. It was
previously shown in section~\ref{prlreview} that increasing shear rates cause the
slip length to decrease. This dependence is related to the reduced height of
the roughness as the shear deforms the bubbles (see also
Ref.~\cite{bib:jens-kunert:2007b}). Similar behavior is found also for the mattress with
bimodal bubble-size distribution (see Fig.~\ref{fig:fixPhi}). However, we
point out that it has recently been shown that depinning of the
contact line at high shear rates may lead to formation of continuous gas film
on the surface, and this depinning transition may change the qualitative
behavior of the shear-rate dependence~\cite{bib:gao-feng:2009}. In the
present work, contact lines are pinned on the hole edges at all shear rates.

To summarize, in this section we simulated laminar flow past bubble arrays consisting of
bubbles of two different sizes, and studied the flow slippage caused by the
bubbles. The slip behavior was found to differ from that previously found
for surfaces made up of evenly sized bubbles. The observed difference is
due to the additional larger-scale roughness that results from the more
complex surface geometry. The presented results indicate that even rather
small non-uniformity in the mattress may considerably increase the slip
and thus complicate the observation of negative slip. On the other hand,
as the negative slip is usually an unwanted property from the application point of
view, non-uniformity and disorder could be utilized to prevent negative
slip and increase the throughput in microfluidic devices. 

\section{Conclusion}
To conclude, we presented lattice Boltzmann simulations of laminar flow
past structured surfaces which are covered with nanobubble arrays. 
After a qualitative comparison to experimentally observed flow over
cylindrical bubbles we demonstrated that our simulations are able to
reproduce the general features predicted by the theoretical description of
Davis and Lauga. However, for a quantitative comparison the theory needs
to be extended to take into account diffuse liquid-gas interfaces as well
as a finite density difference between liquid and gas. By studying arrays
consisting of spherical bubbles we demonstrated that a strong shear can
deform the bubbles and thus reduce the apparent roughness of the surface.
As a consequence the effective slip is reduced. We have further shown that
patterned surfaces can be designed for maximal slippage by variation of
the position of the bubbles as well as by utilizing a distribution of
their size. Our results can be utilized to develop 
micropatterned superhydrophobic surfaces for specific microfluidic applications where a
local variation or an especially high value of the effective slip might be
used to direct flow or to maximize the throughput in a microchannel.
Further, we believe that our results can contribute to the understanding
of some controversially discussed experimental results on the shear rate
dependence of boundary slippage.

\ack
This work was supported by the DFG priority program ``nano- and
microfluidics'' (project Ha~4382/2-1), the Academy of Finland (project
135857) and under the HPC-EUROPA2 project (project 228398). The
computations were performed at the J{\"u}lich Supercomputing Centre,
the Scientific Supercomputing Centre Karlsruhe, and SARA Amsterdam.

\def\newblock{\ }

\end{document}